\documentclass{revtex4}
\usepackage{amsmath}
\usepackage{graphicx}
\usepackage{amssymb}
\usepackage{mathpazo}
\usepackage{float}
\usepackage{dcolumn}

\newcommand{\ignore}[1]{}

\newcommand{\be}{\begin{equation}}
\newcommand{\ee}{\end{equation}}
\def\ba#1\ea{\begin{align}#1\end{align}}
\newcommand{\bit}{\begin{itemize}}
\newcommand{\eit}{\end{itemize}}

\def\slashb#1{\setbox0=\hbox{$#1$}#1\hskip-\wd0\dimen0=5pt\advance
        \dimen0 by-\ht0\advance\dimen0 by\dp0\lower0.5\dimen0\hbox
          to\wd0{\hss\sl/\/\hss}}

\input{epsf}
\input{tcilatex}
\begin{document}

\title{\textbf{Global Analysis of the Source and Detector Nonstandard
Interactions Using the Short Baseline }$\nu -e$\textbf{\ and~}$\overline{\nu 
}-e\ $\textbf{Scattering Data}}
\author{Amir N. Khan}
\email{khan8@mail.sysu.edu.cn \\
ntrnphysics@gmail.com}
\affiliation{School of Physics, Sun Yat-Sen University, Guangzhou, Guangdong 510275, China}

\begin{abstract}
We present a global analysis of the semileptonic and purely leptonic
nonuniversal and flavor-changing nonstandard neutrino interactions in all
the known short-baseline neutrino$-$\ and antineutrino$-$electron scattering
experiments. The nonstandard effects at the source and at the detector can
be more transparent in these experiments because of the negligibly small
ratio between the baselines and the neutrino energies, which is not enough
for the neutrinos to oscillate, and thus can be sensitive to the new physics
at the both ends. We use data from two electron-neutrino electron scattering
experiments and six electron-antineutrino electron scattering experiments
and combine them to find the best fits on the nonstandard parameters using
the source-only, detector-only analyses, and then find the interplay between
the two cases. The bounds obtained in some cases are stronger and new, in
some cases comparable to the current ones, and in the other cases weaker.
For instance, the bound obtained from the interplay between the source and
detector nonstandard physics on the nonstandard parameter $\varepsilon
_{ee}^{udL}$ at the source is much stronger and is comparable with the
indirect bound, but the bounds on the parameters $\varepsilon _{\mu e}^{udL}$
and $\varepsilon _{\tau e}^{udL}$ are weaker in this study in comparison
with the indirect bounds. We also find a global fit on the standard weak
mixing angle $\sin ^{2}\theta _{W}=0.249\pm 0.020$ with $2\%$ improvement in
its precision in comparison with the previous studies.
\end{abstract}

\date{\today }
\maketitle

\section{Introduction}

The discovery of the long-sought neutrino mixing parameter $\theta _{13}$\
in the short-baseline disappearance reactor neutrino oscillation
experiments, Double Chooz, RENO and Daya Bay \cite{DC, DB, RENO} has
completed the list of the unknown mixing parameters of the neutrino
oscillation theory. The goals of the ongoing series of medium-baseline
reactor neutrino disappearance experiments, JUNO and RENO50 \cite{JUNO,
RENO-50}, and the long-baseline accelerator neutrino appearance experiments,
NOvA and DUNE \cite{nova, DUNE}, as well as future neutrino factory
experiments \cite{nufac} are to measure the leptonic CP-violating phase, the
only yet unknown parameter of the theory, which is highly favored for the
matter-antimatter asymmetry of the Universe, to determine the correct
ordering of the neutrino masses, normal or inverted, and to achieve high
precision in the standard mixing parameters and the magnitudes of the
mass-squared differences. The solar neutrino experiment, BOREXINO, has
recently yielded for the first time the real-time measurement of the
low-energy pp neutrinos \cite{Borexino} which, although not very precise,
are a milestone for the solar neutrino flux. Concurrently, LENA is underway
to give a high-precision measurement of the mixing parameters \cite{LENA}.

The short-baseline accelerator neutrino- and reactor antineutrino-electron
elastic scattering experiments \cite{LAMPF, LSND, Irvine, Krasno, Rovno,
MUNU, texono1} have played a key role in confirming the gauge structure of
standard model, in precisely testing the electroweak mixing parameter$,$ $%
\sin ^{2}\theta _{W},$\ \cite{LAMPF, LSND, Irvine, Krasno, Rovno, MUNU,
texono1} and constraining the nonstandard physics parameters \cite%
{Davidson,Forero, JB1,JB2,JB3, ANK2,isodar,texono2}. Though these
experiments are very challenging because of the tiny sizes of cross
sections, they are perfect probes of the precise determination of the $\sin
^{2}\theta _{W}$\ and the nonstandard physics because no complications due
to the hadronic structure are involved. These experiments are also explored
for hints on neutrino magnetic moments \cite{JB3,texono2}, searches for
unparticles \cite{texono2}, neutrino $Z^{^{\prime }}$couplings \cite{Andre},
the large mass-squared difference case due to the sterile neutrinos \cite%
{sterile} ($\sim $1eV$^{\text{2}}$) and the nonstandard neutrino
interactions (NSIs).

The currently running precision neutrino oscillation and neutrino-- and
antineutrino-electron elastic scattering experiments can be used as perfect
probes for the new physics due to the lepton universality violation and the
lepton flavor violation. In the neutrino oscillation experiments,
nonuniversal (NU) and flavor-changing (FC) NSIs are studied in combination
with the neutrino oscillations at the neutrino source, propagation and at
the detector \cite{grossman,ohlsson,biggio,ANK,ozz,gm,gmp,klos,MINOSnsi}. On
the other hand, in the short-baseline neutrino-- and antineutrino-electron
scattering experiments, these NSIs can be studied without the interference
of neutrino oscillations because their baselines and energies result in
small L/E$_{\nu }$ ratios and one can safely ignore the oscillation effects.
Another important aspect of the NSI study in the short-baseline experiments
is that their measurements are independent of the energy resolution of the
detector because the neutrino beam dispersion is negligibly small and there
are no degeneracies between NSIs and the energy uncertainties of the beam 
\cite{Forero,JB1,JB2}. As has been checked in Refs. \cite{Forero,JB1,JB2}
there are no effects of the energy resolution on the size of neutrino cross
sections in these experiments, so we ignore the detector energy resolution
effects for all the reactor neutrino cross sections included in this study.

We use the formalism developed in Ref. \cite{ANK2} and extend the analysis
of reference \cite{Forero}, where only NU NSIs were considered, to find the
global fits on all the possible NSI parameters (NU and FC) at source and
detector and their interplays as we did in Ref. \cite{ANK2} only for the
case of TEXONO experiment. The striking feature of the formalism developed
in Ref. \cite{ANK2} is that it connects the physics at the source and at the
detector. This enables one to explore the semileptonic NSIs at the source
using the recoiled electron data of $\nu -e$\ and $\overline{\nu }-e\ $%
scattering processes. The combination of the $\nu -e$\ scattering data and $%
\overline{\nu }-e\ $scattering data can significantly improve the detector
NSI parameter bounds as the $\nu -e$\ scattering data are sensitive to the
left-handed (LH) couplings, whereas the $\overline{\nu }-e$\ scatterings are
sensitive to the right-handed (RH) couplings. We focus on the leading order
neutrino- and antineutrino interaction processes, calculate their cross
sections in the form appropriate for these experiments and then fit the
global data from these experiments with all the unknown NSI parameters
relevant for the $\nu -e$\ and~$\overline{\nu }-e\ $scatterings.\textit{\ }%
This analysis is based on the combined data of accelerator-based neutrino
sources in the energy range (7-50)\ MeVand the reactor neutrino sources in
the energy range of (3-8) MeV in the respective leptonic scattering
processes.\textit{\ }

In the following section, we review the formalism being used in this
analysis. We then turn to the global fitting analysis for the standard model
parameter, $\sin ^{2}\theta _{W},$\ in Sec. II using the combined data of
the eight scattering experiments. In Sec. III, we find the global fits on
NSI parameters only at the sources, using the recoiled electron energy data;
we call them the source-only parameters. In Sec. IV, we turn to the global
fitting analysis of the NSI parameters only at the neutrino detectors and
them the detector-only parameters, using the recoiled energy spectrum at the
detector. In Sec. V, we find the interplay between the source-only and the
detector-only NSI parameters using the global data. In Sec. VI, we conclude
and summarize.

\section{Formalism and Notations}

\subsection{NSI effective Lagrangians at the source and detector}

For the setup under consideration, the sources of neutrinos are the
charged-current (CC) pion decays at the accelerators and, for antineutrinos,
are the CC neutron beta decays at the reactors, while the target particles
at the detectors are electrons and the contributions for the $\nu -e/%
\overline{\nu }-e$ scatterings come from both the CC and neutral currents
(NC). Therefore, the effective four-fermion LH Lagrangians governing the CC
semileptonic decays at both the accelerators and reactors \cite{jm1,jm2,ANK}
and the LH and RH effective four-fermion Lagrangians for the leptonic $\nu
-e/\overline{\nu }-e$ scattering processes \cite{jm1, jm2, jm3, gb1, gb2} at
the detector are, respectively, given as

\begin{eqnarray}
\mathcal{L}^{s} &=&\mathcal{L}_{NU}^{s}+\mathcal{L}_{FC}^{s} \\
\mathcal{L}^{\ell } &=&\mathcal{L}_{NU}^{\ell }+\mathcal{L}_{FC}^{\ell },
\end{eqnarray}%
where,%
\begin{eqnarray}
\mathcal{L}_{NU}^{s} &=&-2\sqrt{2}G_{F}\sum_{\alpha }(1+\varepsilon _{\alpha
\alpha }^{udL})(\bar{l}_{\alpha }\gamma _{\lambda }P_{L}U_{\alpha a}\nu
_{a})(\bar{d}\gamma ^{\lambda }P_{L}u)^{\dagger }+h.c., \\
\mathcal{L}_{FC}^{s} &=&-2\sqrt{2}G_{F}\sum_{\alpha \neq \beta }\varepsilon
_{\alpha \beta }^{udL}(\bar{l}_{\alpha }\gamma _{\lambda }P_{L}U_{\beta
a}\nu _{a})(\bar{d}\gamma ^{\lambda }P_{L}u)^{\dagger }+h.c., \\
\mathcal{L}_{NU}^{\ell } &=&-2\sqrt{2}G_{F}\sum_{\alpha }(\overline{e}\
\gamma _{\mu }\left( \widetilde{g}_{\alpha R}P_{R}+(\widetilde{g}_{\alpha
L}+1)P_{L})e\right) (\bar{\nu}_{\alpha }\gamma ^{\mu }P_{L}\nu _{\alpha }),
\\
\mathcal{L}_{FC}^{\ell } &=&-2\sqrt{2}G_{F}\sum_{\alpha \neq \beta
}\varepsilon _{\alpha \beta }^{eP}(\bar{e}\gamma _{\lambda }Pe)(\bar{\nu}%
_{\alpha }\gamma ^{\lambda }P_{L}\nu _{\beta }),
\end{eqnarray}

where%
\begin{equation*}
\widetilde{g}_{\alpha R}=\sin ^{2}\theta _{w}+\varepsilon _{\alpha \alpha
}^{eR}\ \text{and}\ \ \widetilde{g}_{\alpha L}=\sin ^{2}\theta _{w}-\frac{1}{%
2}+\varepsilon _{\alpha \alpha }^{eL}.
\end{equation*}%
The superscripts \textquotedblleft \emph{s}\textquotedblright\ and
\textquotedblleft $\emph{l}$\textquotedblright\ designate semileptonic and
leptonic and the subscripts \textquotedblleft NU\textquotedblright\ and
\textquotedblleft FC\textquotedblright\ correspond to the nonuniversal and
flavor-changing NSIs for both the cases, respectively. Here "$\alpha $" and "%
$\beta $" \ are the flavor-basis indices and "$a$" is the mass-basis index
which can be eliminated by the effective replacement of U$_{\alpha a}\nu
_{a}\rightarrow \nu _{\alpha }$\ in Eq. (3) and $U_{\beta a}\nu
_{a}\rightarrow \nu _{\beta }~$in Eq. (4)$,$\ because oscillations play no
role due to the small L/E$_{\nu }$ ratio for these experiments.

For simplicity, we consider only the LH effective Lagrangian at the sources
and ignore the RH part because we do not consider any RH interactions of
neutrinos in this study. The complex coefficients $\varepsilon _{\alpha
\beta }^{udL}$\ represent the relative coupling strengths of the flavor
combinations in the presence of new physics at accelerator or reactor
sources, and the complex coefficients $\varepsilon _{\alpha \beta }^{eP}$\
represent the relative coupling strengths of the flavor combinations in the
presence of new physics at detector to SM, while in the SM case both $%
\varepsilon _{\alpha \beta }^{udL}=0\ $and $\varepsilon _{\alpha \beta
}^{eP}=0.\ $The Hermiticity of the leptonic effective Lagrangian$,$ $L^{\ell
},$\ requires that the detector NSI parameters matrix is Hermitian and
therefore, $\epsilon _{\alpha \beta }^{eR,L}=(\epsilon _{\beta \alpha
}^{eR,L})^{\ast }$, so the NU NSI parameters at the detectors are real, but
the FC NSI parameters are complex in general. With the effective Lagrangians
defined, we are now ready to summarize the cross sections and flux factors
needed for quantifying the NSI effects at the source and detector.

\subsection{$\protect\nu _{e}/\overline{\protect\nu }_{e}-e,\ \protect\nu _{%
\protect\mu }/\overline{\protect\nu }_{\protect\mu }-e$ and $\protect\nu _{%
\protect\tau }/\overline{\protect\nu }_{\protect\tau }-e$ scattering cross
sections}

It is a well-known fact that the\ $\nu _{e}/\overline{\nu }_{e}-e$\
scattering processes get contributions from both the NC and CC interactions$%
,\ $whereas the $\nu _{\mu }/\overline{\nu }_{\mu }-e$\ and $\nu _{\tau }/%
\overline{\nu }_{\tau }-e$\ scattering processes have contributions only
coming from the NC interactions. Therefore, the $\nu _{e}/\overline{\nu }%
_{e}-e$\ scattering cross sections are the coherent sums of the NC and CC
contributions and the $\nu _{\mu }/\overline{\nu }_{\mu }-e$\ and $\nu
_{\tau }/\overline{\nu }_{\tau }-e\ $scattering cross sections have only NC
contributions. All the SM contributions are implicitly given in the
definitions of the parameters $\widetilde{g}_{eL}$\ and $\widetilde{g}_{eR}$%
\ for the $\nu _{e}/\overline{\nu }_{e}-e$\ scattering processes and those
for $\nu _{\mu }/\overline{\nu }_{\mu }-e$\ and $\nu _{\tau }/\overline{\nu }%
_{\tau }-e$\ scattering processes are given in the definitions of $%
\widetilde{g}_{\mu L},\ \widetilde{g}_{\mu R},\ \widetilde{g}_{\tau L}$\ and 
$\widetilde{g}_{\tau R}.$\ The differential cross sections for the three
processes of $\nu _{e}-e,\ \nu _{\mu }-e$\ and $\nu _{\tau }-e$\ scatterings
are given in a compact form as \cite{ANK2},%
\begin{eqnarray}
\left[ \frac{d\sigma (\nu _{\beta }e)}{dT}\right] _{SM+NSI} &=&\frac{%
2G_{F}^{2}m_{e}}{\pi }[\widetilde{g}_{\beta L}^{2}+\underset{\alpha \neq
\beta }{\Sigma }|\varepsilon _{\alpha \beta }^{eL}|^{2}  \notag \\
&&+\left( (\widetilde{g}_{\beta R})^{2}+\underset{\alpha \neq \beta }{\Sigma 
}|\varepsilon _{\alpha \beta }^{eR}|^{2}\right) \left( 1-\frac{T}{E_{\nu }}%
\right) ^{2}  \notag \\
&&-\left( \widetilde{g}_{\beta L}(\widetilde{g}_{eR})+\underset{\alpha \neq
\beta }{\Sigma }\Re \lbrack (\varepsilon _{\alpha \beta }^{eL})^{\ast
}\varepsilon _{\alpha \beta }^{eR}]\right) \frac{m_{e}T}{E_{\nu }^{2}}],
\end{eqnarray}

where $\alpha ,\ \beta =e,\mu ,\tau $ and%
\begin{eqnarray}
\text{ for }\beta &=&e,\ \widetilde{g}_{\beta L}=\widetilde{g}_{eL}\text{ \
and \ \ }\widetilde{g}_{\beta R}=\widetilde{g}_{eR}~+1,~  \notag \\
\text{\ \ for }\beta &=&\mu ,\ \widetilde{g}_{\beta L}=\widetilde{g}_{\mu L}%
\text{ and \ }\widetilde{g}_{\beta R}=\widetilde{g}_{\mu R},\text{\ \ \ } 
\notag \\
\ \text{for }\beta &=&\tau ,\ \widetilde{g}_{\beta L}=\widetilde{g}_{\tau L}%
\text{ \ and \ }\widetilde{g}_{\beta R}=\widetilde{g}_{\tau R},\ \ 
\end{eqnarray}

and the differential cross sections of $\overline{\nu }_{e}-e,\ \overline{%
\nu }_{\mu }-e$\ and $\overline{\nu }_{\tau }-e$\ scatterings are 
\begin{eqnarray}
\left[ \frac{d\sigma (\bar{\nu}_{\beta }e)}{dT}\right] _{SM+NSI} &=&\frac{%
2G_{F}^{2}m_{e}}{\pi }[\widetilde{g}_{\beta R}^{2}+\underset{\alpha \neq
\beta }{\Sigma }|\varepsilon _{\alpha \beta }^{eR}|^{2}  \notag \\
&&+\left( (\widetilde{g}_{\beta L})^{2}+\underset{\alpha \neq \beta }{\Sigma 
}|\varepsilon _{\alpha \beta }^{eL}|^{2}\right) \left( 1-\frac{T}{E_{\nu }}%
\right) ^{2}  \notag \\
&&-\left( \widetilde{g}_{\beta R}(\widetilde{g}_{eL})+\underset{\alpha \neq
\beta }{\Sigma }\Re \lbrack (\varepsilon _{\alpha \beta }^{eR})^{\ast
}\varepsilon _{\alpha \beta }^{eL}]\right) \frac{m_{e}T}{E_{\nu }^{2}}],
\end{eqnarray}

where $\alpha ,\ \beta =e,\mu ,\tau $\ and 
\begin{eqnarray}
\text{for }\beta &=&e,\ \ \widetilde{g}_{\beta L}=\widetilde{g}_{eL}+1,\text{
}\widetilde{g}_{\beta R}=\widetilde{g}_{eR}~~\text{ }  \notag \\
\text{for }\beta &=&\mu ,\ \ \widetilde{g}_{\beta L}=\widetilde{g}_{\mu L},%
\text{ \ \ \ \ \ }\widetilde{g}_{\beta R}=\widetilde{g}_{\mu R}\text{\ \ \ \
\ }  \notag \\
\text{for }\beta &=&\tau ,\ \ \widetilde{g}_{\beta L}=\widetilde{g}_{\tau L},%
\text{ \ \ \ \ \ \ }\widetilde{g}_{\beta R}=\widetilde{g}_{\tau R}\ \ \ 
\end{eqnarray}%
where Eq. $(7)$ and $(9)$ are the sums of the scattering cross sections for
the three incoherent processes corresponding to each index of $\beta .$ For
instance, for $\nu _{e}-e$ scattering, it is the sum of $\nu
_{e}+e\rightarrow \nu _{e}+e,\ \nu _{e}+e\rightarrow \nu _{\mu }+e$\ and $%
\nu _{e}+e\rightarrow \nu _{\tau }+e$ and for $\overline{\nu }_{e}-e$
scattering it is the sum of $\bar{\nu}_{e}+e\rightarrow \bar{\nu}%
_{e}\rightarrow e,\ \bar{\nu}_{e}+e\rightarrow \bar{\nu}_{\mu }+e$\ and $%
\bar{\nu}_{e}+e\rightarrow \bar{\nu}_{\tau }+e$; likewise, for $\nu _{\mu }/%
\bar{\nu}_{\mu }-e$ and $\nu _{\tau }/\bar{\nu}_{\tau }-e$ scattering
processes. Defining the complex parameters $\varepsilon _{\alpha e}^{eL}$\
and $\varepsilon _{\alpha e}^{eR}$\ as $|\varepsilon _{\alpha e}^{eL}|\exp
(i\phi _{\alpha e}^{eL})$\ and $|\varepsilon _{\alpha e}^{eR}|\exp (i\phi
_{\alpha e}^{eR})$, where $\alpha \neq e,\ $and $\phi _{\alpha e}^{eL}$\ and 
$\phi _{\alpha e}^{eR}$\ are the corresponding phases of the complex
quantities, the interference terms in Eq. $(7)$ and Eq. $(9)$ can be written
in a form, that takes into account the phase differences of the NSI
parameters which have been ignored in the previous studies of Refs. \cite%
{Forero,texono2,Davidson,JB1,JB2,isodar}, but were included for the TEXONO
case in Ref. \cite{ANK2}, 
\begin{equation}
\Re \lbrack (\varepsilon _{\alpha e}^{eR})^{\ast }\varepsilon _{\alpha
e}^{eL}]=|\varepsilon _{\alpha e}^{eR}||\varepsilon _{\alpha e}^{eL}|\cos
(\Delta \phi ),  \label{eq:phasefactor}
\end{equation}%
where $\Delta \phi =\phi _{\alpha e}^{eL}-\phi _{\alpha e}^{eR}~$is the
phase difference between the LH and RH FC NSI parameters at the detector.
With this parametrization, the values of $|\varepsilon _{\alpha e}^{eR}|$\
and $|\varepsilon _{\alpha e}^{eL}|$\ are always positive and the sign of
the term is controlled by $\cos (\Delta \phi )$.

\begin{table*}[t]
\begin{center}
\begin{tabular}{l|l|l|l|l|l}
\hline\hline
Experiment & E$_{\nu }(MeV)$ & T(MeV) & Events & Cross-Sections & $\sin
^{2}\theta _{W}$ \\ \hline
LSND \  & $20\ <E_{\nu }<50$ & 20-50 & $191$ & $[10.1\pm 1.86]E_{\nu }\times
10^{-45}$cm$^{2}$ & $0.248\pm 0.051$ \\ 
LAMPF & $7\ <E_{\nu }<50$ & 7-50 & $236$ & $[10.1\pm 1.74]E_{\nu }\times
10^{-45}$cm$^{2}$ & $0.249\pm 0.063$ \\ 
IRVINE\ I & $1.5\ <E_{\nu }<8$ & 1.5-3.0 & $381$ & $[0.87\pm 0.25]\times
\sigma _{V-A}$ & $0.29\pm 0.05$ \\ 
IRVINE\ II\  & $3.0\ <E_{\nu }<8.0$ & 3.0-4.5 & $77$ & $[1.70\pm 0.44]\times
\sigma _{V-A}$ & $0.29\pm 0.05$ \\ 
KRANOYARSK & $3.2\ <E_{\nu }<8.0$ & 3.2-5.2 & N.A & $[4.5\pm 2.4]\times
10^{-46}$cm$^{2}$/fission & $0.22_{-0.8}^{+0.7}$ \\ 
MUNU & $0.7\ <E_{\nu }<8.0$ & 0.7-2.0 & $68$ & $[1.07\pm 0.34]\times $%
events/day & $0.25\pm 0.08^{\ast }$ \\ 
ROVNO & $0.6<E_{\nu }<8.0$ & 0.6-2.0 & $41$ & $[1.26\pm 0.62]\times 10^{-46}$%
cm$^{2}$/fission & $0.29\pm 0.15^{\ast }$ \\ 
TEXONO & $3.0\ <E_{\nu }<8.0$ & 3.0-8.0 & $414\pm 100$ & $[1.08\pm
0.26]\times \sigma _{SM}$ & $0.251\pm 0.04$ \\ \hline
Global & - & - & - & - & $0.249\pm 0.020$ \\ \hline\hline
\end{tabular}%
\end{center}
\caption{List of the accelerator and reactor short basline $\protect\nu -e$
and $\overline{\protect\nu }-e\ $scattering experiments with their energy
ranges (E$_{\protect\nu }$), recoiled electron energies (T), the total
number of observed events, cross-sections and the correponding measured
values $\sin ^{2}\protect\theta _{W}.$ Notice that the entries with * in the
last column are not provided by these experiments, but we find best fits of $%
\sin ^{2}\protect\theta _{W}$ with 1$\protect\sigma $ uncertainty using the
data for MUNU and ROVNO as shown in Fig 2. All of the errors displayed here
are the quadrature sum of the statistical and the systematic uncertainties.
The last row shows the global best fit value of $\sin ^{2}\protect\theta %
_{W} $ with 1$\protect\sigma $ uncertainty.}
\end{table*}

The total cross section for each process will be to integrate over the
recoiled electron energy for the full range of the incoming neutrino beam as
given in Table I. The total cross section for each process then becomes

\begin{eqnarray}
\left[ \sigma (\nu _{\beta }e)\right] _{SM+NSI} &=&\frac{2G_{F}^{2}m_{e}E_{%
\nu }}{\pi }[\widetilde{g}_{\beta R}^{2}+\underset{\alpha \neq \beta }{%
\Sigma }|\varepsilon _{\alpha \beta }^{eR}|^{2}  \notag \\
&&+\frac{1}{3}\left( (\widetilde{g}_{\beta L})^{2}+\underset{\alpha \neq
\beta }{\Sigma }|\varepsilon _{\alpha \beta }^{eL}|^{2}\right)  \notag \\
&&-\left( \widetilde{g}_{\beta R}(\widetilde{g}_{eL})+\underset{\alpha \neq
\beta }{\Sigma }\Re \lbrack (\varepsilon _{\alpha \beta }^{eR})^{\ast
}\varepsilon _{\alpha \beta }^{eL}]\right) \frac{m_{e}}{2E_{\nu }}].
\end{eqnarray}

In the case of antineutrinos, each total cross section is integration over
the recoiled electron energy convoluted by the incoming neutrino spectrum,
energy resolution of the detector, and the efficiency factor; therefore, the
theoretically modeled or expected cross section for each process is,%
\begin{equation}
\left[ \sigma (\overline{\nu }_{\beta }e)\right] _{SM+NSI}=\int_{T^{\min
}}^{T^{\max }}dT\int_{E_{\nu }^{\min }(T)}^{E_{\nu }^{\max }}\frac{d\sigma (%
\bar{\nu}_{\beta }e)}{dT}\times \ \frac{d\ \phi (E_{\nu })}{dE_{\nu }}\
dE_{\nu },
\end{equation}%
where d$\phi $(E$_{\nu }$)/dE$_{\nu }\ $is the reactor antineutrino
spectrum, given as d$\phi $(E$_{\nu }$)/dE$_{\nu }$=$\underset{k=1}{\overset{%
4}{\Sigma }}$a$_{k}\phi _{k}$(E$_{\nu }$),$\ $where a$_{k}$\ are the
abundances of each fission elements, $^{235}$U, $^{239}$Pu,$^{241}$Pu and $%
^{238}$U and $\phi _{k}$(E$_{\nu }$)\ is flux parametrization of each
element and E$_{\nu }^{\min }$(T)=0.5(T+$\sqrt{T^{2}+2m_{e}T}$)\ and E$_{\nu
}^{\max }$(T)=8 MeV \cite{Forero, Pvogal}. Notice that in Eq. (13), we do
not put the efficiency factor explicitly but our calculation must take into
account the efficiency factor where it is required, specially for the MUNU
experiment. As has been checked in Refs. \cite{Forero,JB1,JB2}, there are no
effects of the energy resolution on the size of neutrino cross sections in
the short-baseline experiments, so we ignore the detector energy resolution
effects in this study.

In case of the short-baseline accelerator and reactor antineutrino
scattering experiments, the distance between the source and detector is of
the order of a few tens of meters; therefore, oscillation effects which are
proportional to the factor $\sin ^{2}$(m$_{i}^{2}$-m$_{j}^{2}$)L/4E$_{\nu }$%
, are ignorable in case of the accelerator neutrinos (7MeV$\leq $E$_{\nu
}\leq $50MeV) and the reactor neutrinos (3MeV$\leq $E$_{\nu }\leq $8MeV).
Effectively, the neutrino flavor produced at the accelerator or reactor is
the same as that at the detector. Therefore, only the NSI factors $%
\varepsilon _{\alpha \beta }^{udL}$\ control flux of each neutrino flavor in
the incoming beam to detector. Since the reactor neutrino flux model come
out as a result of a large number of independent nuclear reactions and
accelerator neutrino flux model is the result of a large number of pion
decay reactions, so in the presence of NSIs, the emitted flux can be thought
of as an incoherent sum of $\overline{\nu }_{e},\overline{\nu }_{\mu }$\ and 
$\overline{\nu }_{\tau }$\ with weight factors $|1+\varepsilon
_{ee}^{udL}|^{2},\ |\varepsilon _{e\mu }^{udL}|^{2}$\ and $|\varepsilon
_{e\tau }^{udL}|^{2}$\ and of $\nu _{e},\nu _{\mu }$\ and $\nu _{\tau }$with
the same weights factors as for the case of reactor neutrinos. The source
and detector NSI parameters can, therefore, be combined with each other
through the factor $\mathcal{F}$ as

\begin{equation}
\mathcal{F}(\sin ^{2}\theta _{W},\ \varepsilon _{\alpha \beta }^{udL},\
\varepsilon _{\alpha \beta }^{eR},\ \varepsilon _{\alpha \beta
}^{eL})=\sum_{\alpha =e,\mu ,\tau }|\delta _{e\alpha }+\varepsilon _{e\alpha
}^{udL}|^{2}\left[ \sigma (\nu _{\alpha }e)+\sigma (\overline{\nu }_{\alpha
}e)\right] _{SM+NSI},  \label{F-Fac}
\end{equation}%
which is again the incoherent sum of the three cross sections for the $\nu
_{\alpha }e-$scatterings and the three cross sections for the $\overline{\nu 
}_{\alpha }e-$scatterings as given in Eq. $(12)$ and Eq. $(13)$,
respectively.

\begin{figure}[t]
\begin{center}
\includegraphics[width=3in]{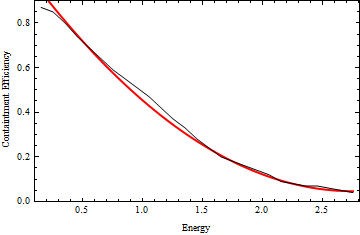}
\end{center}
\caption{Interpolation of the MUNU containment efficiency versus the
antineutrino energy. The black curve is the continuum of the real data of
the MUNU experiment. The red line is the fit to the data \ as defined by the
function : 0.132382 (0.35 + (-2.75 + E$_{\protect\nu }$)$^{2}$)}
\end{figure}

In case of the MUNU experiment, the available data is in the form of event
rates. We have used the real data of the detector's containment efficiency
to calculate the expected event rates, whereas the analysis of Ref. \cite%
{Forero} is based on the assumption of including the normalization in the
theoretical flux as the inverse of the efficiency. The containment
efficiency of the detector versus the antineutrino energy is shown in Fig.
1, where the black curve is the interpolation of the real data points of the
containment efficiency corresponding to the energy range (0.15-2.75) MeV,
whereas red curve is the fit to the data with the fit function, $%
0.132382(0.35+(-2.75+E_{\nu })^{2})$. This function has been used to
calculate the event rate as $1.07\pm 0.042\ $which is consistent with the
one as quoted by the MUNU experiment \cite{MUNU}.

\subsection{The $\protect\chi ^{2}-$fitting model}

To fit the model to the combined data of the accelerator and reactor
experiments and to minimize it for $\sin ^{2}\theta _{W}$ and for the NSI\
parameters, we adopt the $\chi ^{2}$ definition from Refs. \cite%
{Forero,texono2, ANK2,JB1,JB2,JB3} as

\begin{equation}
\chi ^{2}=\sum_{i}\left( \frac{\mathcal{F}_{E}^{i}-\mathcal{F}_{X}^{i}(\sin
^{2}\theta _{W},\varepsilon _{\alpha \beta }^{udL},\varepsilon _{\alpha
\beta }^{eR},\varepsilon _{\alpha \beta }^{eL})}{\Delta ^{i}}\right) ^{2},
\end{equation}%
where $\mathcal{F}_{E}^{i}$\ and\ $\mathcal{F}_{X}^{i}$\ are the
experimental and expected factors which contain the SM and the NSI
contribution of source and detector in the cross sections as defined in Eq. (%
\ref{F-Fac}) for the $i$th\ experiment, and $\Delta ^{i}$ represents the
statistical and the systematic uncertainty of each experiment added in the
quadrature. Since we are using final cross sections and their total
uncertainties from different experiments as the input data points and each
datum is a result of an extensive statistical analysis, the total
uncertainty can be considered as the statistical fluctuation of each datum
in the global analysis like in the case of the standard $\chi ^{2}$
definition used for a single experiment's data points.

We present our analysis for the SM case in terms of $\Delta \chi ^{2}$
versus $\sin ^{2}\theta _{W}$, where $\Delta \chi ^{2}=\chi ^{2}-\chi _{\min
}^{2}$ , for each individual experiment and for the case of global fit.
Similarly, in the case of the minimization of the $\chi ^{2}$ function for
NSI parameters we consider the two-parameter fits using the $\Delta \chi
^{2}\ $versus two NSI parameters each. For this purpose we define $\Delta
\chi ^{2}=\chi ^{2}-\chi _{\min }^{2},$ where $\Delta \chi ^{2}$ is taken to
be 2.71,\ 3.84$\ $and$\ $6.63 corresponding to the $90\%,$ $95\%,$ and $99\%$
confidence levels (C.L.). This is different from Ref. \cite{Forero}, where
the two parameter values of $\Delta \chi ^{2}$ were added with the $\chi
_{\min }^{2}~$to\ get\ their bounds from $1$ parameter projection of the $%
90\%$ C.L. contour boundaries. 
\begin{figure}[t]
\begin{center}
\includegraphics[width=5in]{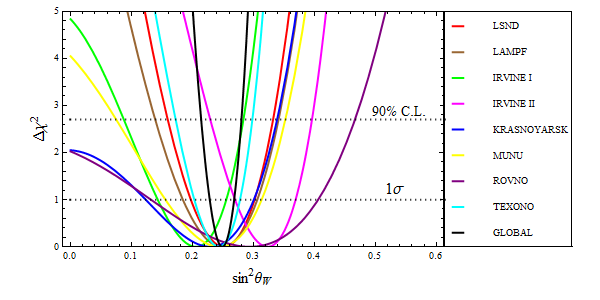}
\end{center}
\caption{{}Best fits of sin$^{2}\protect\theta _{W}$ from the individual
short-baseline accelerator neutrino and reactor antineutrino experiments and
their global $\Delta \protect\chi ^{2}$ fit. Each experimental fit can be
identified by the corresponding legend color as assigned in the vertical
table on the right side. The dotted horizontal lower and upper lines
corresponding to the 1$\protect\sigma $ and 90\% C.L. are shown for
guidance. The global fit corresponds to the black curve. }
\end{figure}

\section{Global Fits of $\sin ^{2}\protect\theta _{W}$, Source and Detector
NSI Parameters Independently}

In this section, we repeat the analysis of Sec. III of Ref. \cite{ANK2}
using the combined data of the reactor- and accelerator -based leptonic
scattering experiments as listed in Table I. We include the FC NSI effects
in the work done in Ref. \cite{Forero}, where only the NU NSIs have been
considered. We further explore the NSI phase effects due the FC NSI
parameters at the detector using the combined data, which has been studied
before in Ref. \cite{ANK2} only for the case of the TEXONO experiment \cite%
{texono1, texono2}.

\subsection{Standard model fits}

Using the\ definition of $\chi ^{2}$ in Eq. $(15)$, we minimize it to obtain
the best fit for $\sin ^{2}\theta _{W}$\ using the combined experimental
data as given in Table I. All the NSI parameters are set equal to zero for
this fitting. We have added IRVIVE II and the TEXONO to the list of
experiments as taken for the same analysis in Ref. \cite{JB3}. The $\chi
^{2}-$minima and the best-fits for each experiment and the global minimum
and the related best fit value are all shown in Fig. 2.\ Each experimental
input is identified by a particular color assigned in the legend aside. The
black curve shows the global best fit with a value of $\sin ^{2}\theta
_{W}=0.249\pm 0.020$\ at $\ $minimum-$\chi ^{2}$\ of $2.68$. The uncertainty
shown is $1\sigma $\ when the minimum-$\Delta \chi ^{2}=0$, a better fit
than the one obtained in Ref. \cite{JB3} having precision of $10\%$ at $%
1\sigma ,$ whereas in our case we have precision of $8\%$.\ The 1$\sigma $\
and $90\%$\ C.L. lines are included for guidance in Fig. $2$.

\subsection{NSIs with the Source-Only Case}

Here we present the analysis for the source-only (semileptonic) NSI
parameters using the combined data of the accelerator and reactor
experiments as listed in Table I. As discussed in Sec. I, although we are
using the data from the recoiled electron energy spectrum at the detector,
our formalism still allows us to obtain constraints on the source-only NSI
parameters. We take contour boundaries of the source-only NSI parameters in
the $($NU$(\varepsilon _{ee}^{udL})-$FC$(\varepsilon _{\alpha e}^{udL}))$
parameter space and then extract bounds from the projections on the relevant
axes as shown in Fig. $3$. The figure shows the boundaries of the
two-parameter fits to the combined data of the experiments at a $90\%,\
95\%, $ and $99\%\ $C.L. boundaries as colored as red (inner), blue (middle)
and green (outer), respectively. The bounds obtained from the contours
corresponding to the $90\%~$C.L. are given in the first two rows of Table II.

\begin{figure}[t]
\begin{center}
\includegraphics[width=2.5in]{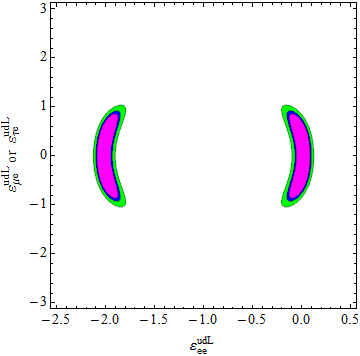}
\end{center}
\caption{Global analysis for the source-only NSI parameters case. The three
different contour regions correspond to the $90\%,\ 95\%$ and $99\%$ C.L.
from inner to outer, respectively.}
\end{figure}

\subsection{NSIs with the Detector-Only Case and the NSI Phase Effects}

A similar exercise, as in Sec. IIIB, is repeated for the detector-only
(leptonic) NSI parameters and the results are shown in Fig. 4, where Fig.
4(a) shows the contour boundaries of the detector-only NU NSI parameters and
the other three panels show the contours of the detector FC NSI parameters
at $90\%,\ 95\%,$ and $99\%$ C.L.. Figs. (4(b)-4(d)) show effects of the NSI
phase, appearing in the RH-LH interference term of Eq. (7) and Eq.\ (9) on
the C.L. boundaries, when the NSI phases have values $\cos (\Delta \phi )$\
= $1,0$ and $-1$, which is coming from the RH-LH interference terms of the
FC NSI parameters at the detector in the differential cross sections. Each
choice of the phase part corresponds to the different choices of the NSI
phases and their differences. For example, Fig. 4(d) corresponds to the
composite of the cases $\phi _{\alpha e}^{eR}=\phi _{\alpha e}^{eL}=0$,
where $\epsilon _{\alpha e}^{eR}$\ and $\epsilon _{\alpha e}^{eL}$\ are both
real and positive, $\phi _{\alpha e}^{eR}=\pi $\ and $\phi _{\alpha
e}^{eL}=0 $, where $\epsilon _{\alpha e}^{eR}$\ is real and negative and $%
\epsilon _{\alpha e}^{eL}$\ is real and positive, $\phi _{\alpha
e}^{eR}=\phi _{\alpha e}^{eL}=\pi $, where $\epsilon _{\alpha e}^{eR}$\ and $%
\epsilon _{\alpha e}^{eL}$\ are both real and negative, and, finally, $\phi
_{\alpha e}^{eR}=0$\ and $\phi _{\alpha e}^{eL}=\pi $, where $\epsilon
_{\alpha e}^{eR}$\ is real and positive and $\epsilon _{\alpha e}^{eL}$\ is
real and negative. Alternatively, it can be interpreted as the composite of
cases where $0$ and $\pi $\ are replaced with $\pi /2$\ and $3\pi /2$\ and
the real is replaced with the imaginary.\ Similarly, Fig. 4(c) corresponds
to those composite choices in which $\phi _{\alpha e}^{eL}-\phi _{\alpha
e}^{eR}=\pi /2.$\ For such choices the correlation between the RH and LH FC
NSI parameters appearing the LH-RH interference term disappears, because RH
and LH parameters are $\pi /2$\ out of phase; for example, one can be real
and the other can be imaginary.

It is clear from Eq. (7) and (9), the RH-LH interference term of the FC\ NSI
parameters are suppressed by the factor $m_{e}T/E_{\nu }^{2}$, where the
mean of the lower end of the neutrino energies listed in the Table I is
greater than 6.5 MeV and effects of the phases are thus very small, as shown
in Figs. 4(b), 4(c) and 4(d). Conclusions about allowed boundaries for NSI
parameters for the range of energies of interest in the combined accelerator
and reactor short-baseline experiment are affected very little in this
analysis, but for experiments with significantly lower energy radioactive
sources or for low-energy solar neutrinos such as those coming from pp, $%
^{7} $Be, and B$^{8}\ $reactions, and marginally for the lower end of the
pep spectrum, which are all measured in Gallium \cite{SAGE} and BOREXINO 
\cite{borex} experiments, the RH-LH correlation term can be relatively
larger and the phase effects must become important.

\begin{figure}[t]
\begin{center}
\includegraphics[width=7in]{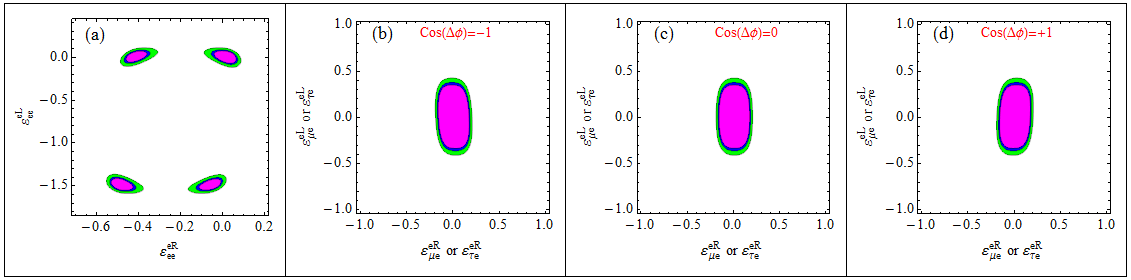}
\end{center}
\caption{Global analysis for the detector-only NSI parameters case. In each
panel, the three different contour regions correspond to the $90\%$, $95\%,$
and $99\%$ C.L. from inner to outer, respectively. Panel (a) shows the NU
NSI\ parameter boundary regions, whereas (b) corresponds to $\cos (\Delta 
\protect\phi )=-1$, (c) to $\cos (\Delta \protect\phi )=0$ and (d)
corresponds to $\cos (\Delta \protect\phi )=+1$ of the FC\ NSI parameter
boundary regions. }
\end{figure}

\begin{table*}[t]
\begin{center}
\begin{tabular}{l|l|l}
\hline\hline
Figure No. & RH-Parameter Bounds & LH-Parameter Bounds \\ \hline
$3$ & $-$ & $-0.13<\varepsilon _{ee}^{udL}<0.10$ \\ 
$3$ & $-$ & $-0.84<\varepsilon _{\alpha e}^{udL}<0.84$ \\ 
$4(a)$ & $-0.04<\varepsilon _{ee}^{eR}<0.06$ & $-0.08<\varepsilon
_{ee}^{eL}<0.08$ \\ 
$4(b),\cos (\Delta \phi )=-1$ & $-0.17<\varepsilon _{\alpha e}^{eR}<0.18$ & $%
-0.33<\varepsilon _{\alpha e}^{eL}<0.35$ \\ 
$4(c),\cos (\Delta \phi )=0$ & $-0.15<\varepsilon _{\alpha e}^{eR}<0.16$ & $%
-0.33<\varepsilon _{\alpha e}^{eL}<0.35$ \\ 
$4(d),\cos (\Delta \phi )=+1$ & $-0.16<\varepsilon _{\alpha e}^{eR}<0.17$ & $%
-0.33<\varepsilon _{\alpha e}^{eL}<0.35$ \\ \hline\hline
\end{tabular}%
\end{center}
\caption{{}{}Bounds at $90\%\ $C.L. obtained from Fig. 3 for the source-only
and from Fig. 4 for the detector-only case where $\protect\alpha =\protect%
\mu ~$or$\ \protect\tau .$ }
\end{table*}

\section{Correlation between the source$\ $and\ detector NSI parameters}

This section is a recap of section IV of Ref. \cite{ANK2} using the combined
data of the short-baseline accelerator and reactor $\nu _{e}-e$\ and $%
\overline{\nu }_{e}-e$\ scattering experiments. Here the joint two-
parameters C.L. boundary regions are taken where one is the \ source
semileptonic NSI parameter and the other is the detector leptonic NSI
parameter. First, we check the boundaries for the NU source versus all the
detector NSI parameters and then FC source versus all the detector NSI
parameters. Bounds are extracted for each parameter using its projection on
the corresponding axis at $90\%$ C.L..

\subsection{The NU source ($\protect\varepsilon _{ee}^{udL})$ versus all the
detector ($\protect\epsilon _{\protect\alpha \protect\beta }^{eR,\text{ }L}$%
) NSI cases}

\begin{figure}[t]
\begin{center}
\includegraphics[width=7in]{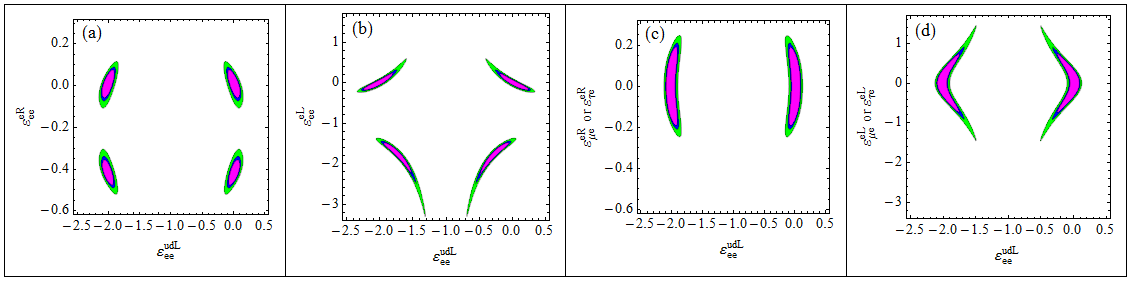}
\end{center}
\caption{Global analysis for the interplay between the source and detector
NSI parameters. The three different contour regions in each panel correspond
to the $90\%,\ 95\%$ and $99\%$ C.L. from inner to outer, respectively.}
\label{Keps1}
\end{figure}

We take pairs of NU source and all of the detector NSI parameters to survey
the $90\%,\ 95\%,$ and the $99\%$ C.L. boundaries in the two-parameter
spaces. Only bounds at $90\%$ C.L. are extracted and are displayed in Table
III. The results of this analysis are displayed in Fig. \ref{Keps1} showing
the C.L. boundaries for the fits to the combined data of the accelerator and
reactor short-baseline scattering data parametrized by one source NSI
parameter and one detector parameter with all of the other NSI parameters
set to zero. From Fig. \ref{Keps1}, we can determine the $90\%$ C.L. bounds
on the source NU parameter$,$ $\varepsilon _{ee}^{udL},\ $and on any of the
detector NSI parameters$,$ $\epsilon _{\alpha e}^{eR,L},$ where $\alpha =e,\
\mu ,$ $\tau $, by projecting onto the parameter axis for each contour. All
of the extracted limits are given in Table \ref{tabKeps1}.

\begin{table*}[tbph]
\begin{center}
\begin{tabular}{l|l|l}
\hline\hline
Figure No. & NSI Parameters at Source & \ NSI\ Parameters at Detector \\ 
\hline
$5(a)$ & $-0.09<\varepsilon _{ee}^{udL}<0.10$ & $-0.06<\varepsilon
_{ee}^{eR}<0.07$ \\ 
$5(b)$ & $-0.19\ \ <\varepsilon _{ee}^{udL}~<0.25$ & $-0.17\ <\varepsilon
_{ee}^{eL}\ <0.23$ \\ 
$5(c)$ & $-0.09<\varepsilon _{ee}^{udL}\ <0.09\ $ & $-0.19<\varepsilon
_{\alpha e}^{eR}<0.19$ \\ 
$5(d)$ & $-0.25<\varepsilon _{ee}^{udL}<0.09$ & $-0.73<\varepsilon _{\alpha
e}^{eL}<0.75$ \\ \hline\hline
\end{tabular}%
\end{center}
\caption{Bounds at $90\%\ $C.L. obtained from Fig. $5$ for the NU source ($%
\protect\varepsilon _{ee}^{udL})$ versus all the detector ($\protect\epsilon %
_{\protect\alpha \protect\beta }^{eR,\text{ }L}$) NSIs case where $\protect%
\alpha ,\protect\beta =\protect\mu ,\protect\tau .$}
\label{tabKeps1}
\end{table*}

\begin{figure}[tbph]
\begin{center}
\includegraphics[width=7in]{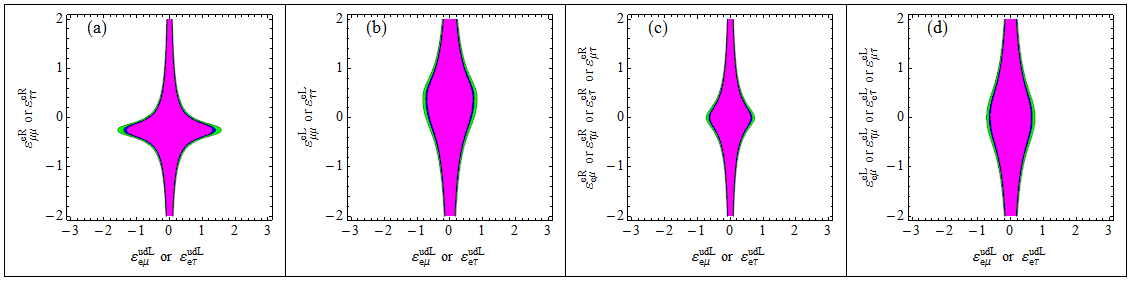}
\end{center}
\caption{{}C.L. boundaries for the correlation between the source FC NSI and
all the detector (NU and FC) NSI parameters using the combined data of the
very short baseline accelerator and reactor neutrino data at $90\%$, $95\%$
and $99\%$ C.L. The red, blue and green colors correspond to the three C.L.
regions, respectively. }
\end{figure}

\begin{table*}[tbph]
\begin{center}
\begin{tabular}{l|l|l}
\hline\hline
Figure no. & NSI parameters at Source & \ NSI\ parameters at detector \\ 
\hline
$6(a)$ & $-1.3\ \ <\varepsilon _{\alpha e}^{udL}\ <1.3$ & $unbounded$ \\ 
$6(b)$ & $-0.70<\varepsilon _{\alpha e}^{udL}\ <0.70$ & $unbounded$ \\ 
$6(c)$ & $-0.62\ <\varepsilon _{\alpha e}^{udL}<0.62$ & $unbounded$ \\ 
$6(d)$ & $-0.62\ <\varepsilon _{\alpha e}^{udL}\ <0.62$ & $unbounded$ \\ 
\hline\hline
\end{tabular}%
\end{center}
\caption{Bounds at $90\%\ $C.L. obtained from Fig. $(6)$ for the FC source ($%
\protect\varepsilon _{e\protect\mu }^{udL}$ or $\protect\varepsilon _{e%
\protect\tau }^{udL})$ versus all the detector ($\protect\epsilon _{\protect%
\alpha \protect\beta }^{eR,\text{ }L}$) NSI cases where $\protect\alpha =%
\protect\mu ,$ $\protect\tau .$}
\label{SourceFC1}
\end{table*}

\subsection{The FC source ($\protect\varepsilon _{e\protect\mu }^{udL}$ or $%
\protect\varepsilon _{e\protect\tau }^{udL})$ versus all the detector ($%
\protect\epsilon _{\protect\alpha \protect\beta }^{eR,\text{ }L}$) NSI cases}

Here we present the interplay between the source FC NSI parameters and all
of the corresponding detector NSI parameters (both NU and FC) and find the
C.L. boundary regions at $90\%,\ 95\%,$ and $99\%$ C.L. as shown in Fig. $6$%
. The bounds are extracted from the 90\% C.L. boundaries and are given in
Table \ref{SourceFC1}.\ In this case only the bounds on the source NSI
parameters$,\ \varepsilon _{e\mu }^{udL},$\ can be extracted while detector
NSI parameters$,\ \epsilon _{\alpha \beta }^{eR,L},$ remain unbounded. This
is because the source is receiving $\bar{\nu}_{e}$ flux in the limit when $%
\varepsilon _{\alpha \beta }^{udL}\rightarrow 0$.$\ $This shows that the
source NSI parameters$,$ $\varepsilon _{\alpha \beta }^{udL},$\ and detector
NSI parameters$,$ $\epsilon _{\alpha \beta }^{eR,L},\ $are highly
correlated. One can see that there is still a possibility for placing upper
bounds on source NSI parameters$,$ $\varepsilon _{e\alpha }^{udL},$ at $90\%$
C.L. when the detector NSI parameters$,$ $\epsilon _{\mu \mu }^{eR,\ L}$and $%
\epsilon _{\alpha \mu }^{eR,\ L},\ $are zero , and likewise for $\mu
\rightarrow \tau $. These are the so-called one-parameter-at-a-time bounds
on the source NSI parameters commonly reported in the literature. The
possible one-parameter-at-a-time bounds are given as asterisk entries in the
3rd column of Table V. We can see from Fig. $6$ that in the case of FC
source NSI parameters, there is no dependence on the detector NSI phase,
because the phase effects are coming from the interference between the LH
and RH parameters of the detector NSIs, but in this case the LH and RH
parameters' contribution is coming separately in the two-parameters
analysis; therefore, these results are not sensitive to the detector NSI
phases.

Summarizing this section we can notice from the Tables \ref{tabKeps1} and %
\ref{SourceFC1} that the bounds obtained from the two-parameter analysis of
the same handedness are weaker, but the bounds with opposite handedness are
stronger.

\section{Summary and Conclusions}

We have extended the work of Refs. \cite{JB3,Forero} by applying the
formalism developed in Ref. \cite{ANK2} using the data of all the available
accelerator and reactor neutrino scattering experiments, as listed in Table
I, to find the global fit of the weak mixing parameter$,$ $\sin ^{2}\theta
_{W},$ and to constrain the NU and FC NSI parameters, along with the NSI
phase effects at the detector. The formalism developed in Ref. \cite{ANK2}
can combine the NSI effects in the semileptonic decays at the accelerator
and reactor neutrinos sources with the leptonic NSI effects at the detector.
This formalism helps to find the sensitivity of the source NSIs using the
detector data of the recoiled electrons. In addition, we can find the
correlation between the source and detector NSI parameters.

We have first used the leptonic scattering data of the accelerator and
reactor experiments to find the best-fit value of $\sin ^{2}\theta _{W},$ as
done before in Ref. \cite{JB3}, with the additional data set from the recent
TEXONO experiment \cite{texono1} added here. These combined data were then
used to explore the NSIs at source and at the detector. We used the
two-parameter joint C.L. boundaries for constraining the source and detector
NSI parameters separately and then found the interplay between them.

In the case of\ $\sin ^{2}\theta _{W}$, we find the minimum-$\chi ^{2}$ of $%
\chi _{\min }^{2}/$d.o.f$=2.68/7$. The results are consistent with the
analysis when one experiment is excluded and the whole exercise is repeated
for one experiment fewer than the total. We find a global best fit value of $%
\sin ^{2}\theta _{W}=0.249\pm 0.020$\ at $1\sigma $ with $8\%$ precision,
which is a $2\%$ improvement in\ the previous value of $\sin ^{2}\theta
_{W}=0.259\pm 0.025$\ at $\chi _{\min }^{2}/$d.o.f$=2.17/6$ of Ref. \cite%
{JB3}.

For the NSI case, we first used the model for the combined data when only
the source semileptonic NSI parameters were considered while all the
detector NSIs were set to zero. The joint two-parameter analysis between the
source NU and the source FC NSIs were performed as shown in Fig. 3 and the
corresponding bounds were obtained as given in Table II, and in the first
three entries of Table V for comparison with previous bounds of Ref. \cite%
{JB1,JB2} and with the indirect bounds of Ref. \cite{biggio}. As clear from
Table V, the bound on the NU source parameter $,\ \varepsilon _{ee}^{udL},$
in the case of the correlated analysis is much stronger and is comparable
with the bounds from the indirect study of the fourth column in the table.
Similarly the source FC NSI parameters$,$ $\varepsilon _{\mu e}^{udL}$ and $%
\varepsilon _{\tau e}^{udL},$ are weaker in this study in comparison with
the indirect bounds, but are new in the sense that these have been obtained
using the recoiled electron data at the detector. As shown in Ref. \cite%
{ANK2}, improvement in the statistical uncertainty of the TEXONO experiment
and use of this framework can improve the source NSI\ bounds in the
scattering experiments. Similarly, the bounds obtained from the global data
can also be improved.

In the case of detector-only NSIs, we performed the two-parameter best-fit
analysis and have shown our results in Fig. 4 and the bounds obtained in
Table II. The LH NU NSI parameter$,$ $\varepsilon _{ee}^{eL},~$at the
detector has a new stringent lower bounds with an order of magnitude
improvement from the one previously obtained using a similar analysis. Since
the FC parameters at the detector also contain the NSI phase contributions,
we show them in three different panels in Fig. 4. Although the NSI phases do
not have significant effects on the boundaries of the NSI parameters space,
however they could have significant effects in the low energy experiments as
has been shown in Ref. \cite{ANK2} in the section of future prospect study
of TEXONO experiment.

In the last section, a correlation between the source semileptonic and the
detector purely leptonic NSI parameters has been explored. The C.L.
boundaries and the bounds obtained from them are given in Figs. $5\ $and $6$
and in Tables III and IV. In Sec. IVA, we explore the interplay between the
source NU NSI parameter$,$ $\varepsilon _{ee}^{udL},$ versus all the
detector NSI parameters$,$ $\epsilon _{\alpha \beta }^{eR,\text{ }L},$
whereas Sec. IVB deals with the interplay between the source FC NSIs$,$ $%
\varepsilon _{e\mu }^{udL}$ or $\varepsilon _{e\tau }^{udL},$ versus all the
detector NSI parameters$,$ $\epsilon _{\alpha \beta }^{eR,\text{ }L}$.

The best amid the bounds in this study are summarized in Table V along with
the previous bounds obtained using the similar processes of $\nu _{e}-e$ and 
$\overline{\nu }_{e}-e\ $scatterings in Refs. \cite{Davidson,Forero,
JB1,JB2,JB3, ANK2, texono2} and with the indirect bounds of Ref. \cite%
{biggio} for comparison. 
\begin{table*}[t]
\begin{center}
\begin{tabular}{l|l|l|l|l}
\hline\hline
NSI parameters & \ This work (uncorrelated) & This work (correlated) & 
Previous bounds & M.I.Bs \\ \hline
\multicolumn{1}{c|}{$\varepsilon _{ee}^{udL}$} & $-0.13<\varepsilon
_{ee}^{udL}<0.10$ & $-0.09<\varepsilon _{ee}^{udL}<0.09$ & 
\multicolumn{1}{|c|}{--} & \multicolumn{1}{|c}{$<0.042$} \\ 
\multicolumn{1}{c|}{$\varepsilon _{\mu e}^{udL}$} & $-0.84<\varepsilon _{\mu
e}^{udL}<0.84$ & $-0.62<\varepsilon _{\mu e}^{udL}<0.62^{\ast }$ & 
\multicolumn{1}{|c|}{--} & \multicolumn{1}{|c}{$<0.042\ $} \\ 
\multicolumn{1}{c|}{$\varepsilon _{\tau e}^{udL}$} & $-0.84<\varepsilon
_{\tau e}^{udL}<0.84$ & $-0.62<\varepsilon _{\tau e}^{udL}<0.62^{\ast }$ & 
\multicolumn{1}{|c|}{--} & \multicolumn{1}{|c}{$<0.042$} \\ 
\multicolumn{1}{c|}{$\varepsilon _{ee}^{eL}$} & $-0.08<\varepsilon
_{ee}^{eL}<0.08,\ $ & $-0.17<\varepsilon _{ee}^{eL}<0.23$ & $%
-0.13<\varepsilon _{ee}^{eL}<0.12$ & \multicolumn{1}{|c}{$<0.06$} \\ 
\multicolumn{1}{c|}{$\varepsilon _{ee}^{eR}$} & $-0.04<\varepsilon
_{ee}^{eR}<0.06$ & $-0.06<\varepsilon _{ee}^{eR}<0.07$ & $-0.07<\varepsilon
_{ee}^{eR}<0.15$ & \multicolumn{1}{|c}{$<0.14$} \\ 
\multicolumn{1}{c|}{$\varepsilon _{\mu e}^{eL}$} & $-0.33<\varepsilon _{\mu
e}^{eL}<0.35$ & $-0.73<\varepsilon _{\mu e}^{eL}<0.75$ & $\ \
-0.43<\varepsilon _{\mu e}^{eL}<0.43$ & \multicolumn{1}{|c}{$<0.10$} \\ 
\multicolumn{1}{c|}{$\varepsilon _{\mu e}^{eR}$} & $-0.15<\varepsilon _{\mu
e}^{eR}<0.16$ & $-0.19<\varepsilon _{\mu e}^{eR}<0.19$ & $\ \
-0.31<\varepsilon _{\mu e}^{eR}<0.31$ & \multicolumn{1}{|c}{$<0.10$} \\ 
\multicolumn{1}{c|}{$\varepsilon _{\tau e}^{eL}$} & $-0.33<\varepsilon
_{\tau e}^{eL}<0.35$ & $-0.73<\varepsilon _{\tau e}^{eL}<0.75$ & $%
-0.43<\varepsilon _{\tau e}^{eL}<0.43$ & \multicolumn{1}{|c}{$<0.40$} \\ 
\multicolumn{1}{c|}{$\varepsilon _{\tau e}^{eR}$} & $-0.15<\varepsilon
_{\tau e}^{eR}<0.16$ & $-0.19<\varepsilon _{\tau e}^{eR}<0.19$ & $\ \
-0.31<\varepsilon _{\tau e}^{eR}<0.31$ & \multicolumn{1}{|c}{$<0.27$} \\ 
\hline\hline
\end{tabular}%
\end{center}
\caption{Bounds for comparison of the model independent study of Ref. 
\protect\cite{biggio}. The asterisk entries correspond to the
one-parameter-at-a-time bounds. Bounds of fourth column with the title
"previous bounds" have been taken from Refs. \protect\cite%
{JB1,JB2,JB3,Forero}, which uses analysis similar to this work. M.I.Bs refer
to the model independent or indirect bounds and have been taken from Ref. 
\protect\cite{biggio}. The "uncorrelated" refers to the bounds taken from
the best among the detector-only and the source-only analyses, while
"correlated" refers to the bounds taken from the best among the combined
analysis of interplay between the source and detector NSI parameters.}
\end{table*}

\begin{acknowledgments}
I am very grateful to Professor Douglas McKay of University of Kansas for
his continuous encouragement and comments during the course of this work. I
also thank Dr. David Forero for the detailed discussions and Dr. Jean-Luc
Vuilleumier of MUNU Collaboration for sharing the information and details
about the MUNU experiment. This work has been partially supported by the Sun
Yat-Sen University under the Post-Doctoral Fellowship program.
\end{acknowledgments}

\end{document}